\documentclass[twocolumn, secnumarabic, amssymb, nobibnotes, aps]{revtex4-1}
\usepackage{amsmath,amstext,bm,appendix,graphicx,multirow,amsfonts}
\usepackage{color}
\usepackage{diagbox}
\usepackage[bookmarks=false]{hyperref}
\hypersetup{colorlinks=true,citecolor=blue,
linkcolor=blue,urlcolor=blue,pdfstartview=FitH,bookmarksopen=true}

\begin{document}
\title{Universal speeded-up adiabatic geometric quantum computation in three-level systems via counterdiabatic driving}
\author{J. L. Wu}
\affiliation{School of Physics and Engineering, Zhengzhou University, Zhengzhou 450001, China}
\author{S. L. Su}
\email[]{slsu@zzu.edu.cn}
\affiliation{School of Physics and Engineering, Zhengzhou University, Zhengzhou 450001, China}

\begin{abstract}
Universal speeded-up adiabatic geometric quantum computation~(SAGQC) is studied in $\Lambda$-type three-level system with different coupling cases, i.e., time-dependent detuning, large detuning and one-photon resonance couplings, respectively. In these cases, the counterdiabatic driving method is used to speed up the universal quantum computation. These schemes in $\Lambda$-type three-level system are feasible in experiment because additional unaccessible ground-state coupling is not needed. Only the shapes and phases of the initial adiabatic classical-field pulse are modified with the aid of effective two-level systems based on the counterdiabatic driving. The speed and robustness against decay of these schemes are discussed and compared. In addition, our work enriches the study of the speeded-up geometric computation in $\Lambda$-type three-level system and can be applied to experimental platforms with different coupling features.
\end{abstract}
\maketitle

\section{Introduction}\label{S1}
Quantum computers are of more capacity and efficiency than their classical counterparts because of their strong power for quantum algorithms~\cite{Shor1997,Grover1997} and parallel computations~\cite{Gingrich2000,Paredes2005,Alvarez2008}. A universal set of high-fidelity quantum gates of accurate quantum computations is indispensable for the physical implementation of a quantum computer. In practice, however, the realization of quantum computer is of great challenge due to the presence of unavoidable noises induced by the interaction between the computational system and its environment, and errors from imprecise manipulations of the computational system. The introduction of geometric phases~\cite{Berry1984,Wilczek1984} into quantum computation is a promising way to suppress control errors~\cite{Zanardi1999}, since geometric phases do not depend on dynamical details but only paths of the quantum-system evolution, which makes quantum computation based on geometric phases insensitive to control errors~\cite{Zhu2005,Wu2005,Xu2012,Berger2013}. In the past several years, many experiments were devoted to the realization of geometric quantum computation in some physical systems including superconducting circuit systems~\cite{Abdumalikov2013,Wang2018NJP,Xu2018PRL,Yan2018}, NMR systems~\cite{Jones2000,Feng2013PRL} and diamond nitrogen-vacancy~(NV) center systems~\cite{AC2014NC,Zu2014,Zhou2017PRL,FK2018}.

Under the regime of adiabatic evolution, Duan \emph{et al.} proposed a four-level scheme of the universal adiabatic geometric quantum computation~(AGQC) in a trapped-ion system~\cite{Duan2001}. However, adiabatic evolution requiring very-long evolution time usually reduces execution efficiency and enhances decoherence accumulation, and thus destroys the desired dynamics. Subsequently, therefore much effort was made on nonadiabatic geometric quantum computation~\cite{Xu2012,Sjoqvist2012,Liang2014,GFXu2015,ZYXue2017,PZZhao2018} which enables high-speed quantum gate operations and thus prevents quantum computation from environment-induced decoherence to some extent. Nonetheless, additional fluctuating phase shifts would be introduced and then disturb the desired geometric phases~\cite{Thomas2011}.

Alternatively, to shorten the evolution time but hold the robustness of the adiabatic evolution, the concept of ``shortcuts to adiabaticity"~(STA) has been proposed~\cite{Berry2009,XiChen2010,Torrontegui2013}. Methods of STA, including Lewis-Riesenfeld invariants~\cite{Lewis1969,XiChen}, transitionless quantum driving~(TQD)~\cite{Berry2009,XiChen2010,Muga2010}, superadiabatic iteration~\cite{Berry1987,Ibanez,XKSong2016}, dressed-state-based inverse engineering~\cite{Baksic2016,Kang2016SR,Wu2017OE} and others~\cite{Opatrny2014,YHChen,YCLi2018}, have been proposed to speed up adiabatic evolution in recent years, and also experimental realizations have been achieved in different physical systems, such as optical-lattice system~\cite{Bason2012}, NV center systems~\cite{Zhang2013,BBZhou2017} and cold-atom system~\cite{YXDu2016}. Up to now, schemes have been proposed to speed up the implementation of AGQC~\cite{JZhang2015SR,Pyshkin2015,XKSong2016NJP,ZTLiang2016,BJLiu2017}. Using TQD, Zhang~\emph{et al.}~\cite{JZhang2015SR} generalized TQD to speed up an implementation of adiabatic non-Abelian~\cite{Wilczek1984} geometric gates~(i.e., holonomic gates~\cite{Zanardi1999}); Song~\emph{et al.}~\cite{XKSong2016NJP} and Liang~\emph{et al.}~\cite{ZTLiang2016} put forward two proposals for accelerating universal holonomic and Abelian~\cite{Berry1984} AGQC in NV centers, respectively. Using dressed-state-based inverse engineering, Liu~\emph{et al.}~\cite{BJLiu2017} proposed a scheme to implement superadiabatic holonomic quantum computation in cavity QED. With the great development of STA in inverse engineering and optimized control~\cite{Ruschhaupt2012,YDWang2017,Mortensen2018}, even Kang~\emph{et al.}~\cite{YHKang2018} constructed nonadiabatic two-qubit holonomic gates with two Rydberg atoms.

The schemes~\cite{JZhang2015SR,XKSong2016NJP,BJLiu2017} of universal non-Abelian SAGQC require quantum systems with at least four levels. The scheme~\cite{ZTLiang2016} of universal Abelian SAGQC is based on a two-level quantum system. Obviously, three-level systems are absent in implementing universal SAGQC. Three-level systems, especially with two lower-energy levels~($\Lambda$-type systems), are promising to implement robust quantum computation, because qubits encoded on the two lower-energy levels are of great stability. Comparing with three-level systems, four-level systems require more complicated level structure and more external driving, and two-level systems are usually unstable for implementing quantum computation because the easily-dissipative higher-energy level has to be used to encode qubits. In a recent study, Liu~\emph{et al.}~\cite{BJLiu2018} proposed to implement SAGQC in a three-level system with time-dependent one-photon detuning $\Delta(t)$ by using the method of Lewis-Riesenfeld invariants.

In this work, counterdiabatic driving~(equivalent to TQD)~\cite{DR20082003,Berry2009,Campo2013}, adding a suitable ``counterdiabatic term" into
an original time-dependent Hamiltonian to suppress transitions between different instantaneous eigenstates, is applied to speed up universal AGQC in three-level systems, including three cases, i.e., time-dependent intermediate-level detuning, large intermediate-level detuning and one-photon resonance. With the aid of effective two-level systems, counterdiabatic driving is experimentally feasible because no additional unaccessible coupling is introduced but only the classical-field pulse shapes and phases are modified in all the three cases. In addition to the advantages of SAGQC, this work has following advantages: (i) The work enriches the investigations of $\Lambda$-type three-level system in implementing SAGQC. (ii) The schemes of SAGQC based on different coupling conditions provides more or better choices for practical realization of universal geometric quantum computation. (iii) The implementation of the proposals does not rely on a certain specific physical system, any experimental platform that has $\Lambda$-type three-level configuration can be used to implement the scheme.

\section{Universal single-qubit gates~(USQG)}\label{S2}
Consider a general two-photon-resonance $\Lambda$-type three-level system with one higher-energy states $|e\rangle$ and two lower-energy states $|0\rangle$ and $|1\rangle$. Under the regime of the stimulated Raman adiabatic passage~(STIRAP) within the rotating wave approximation, the Hamiltonian with the basis states \{$|0\rangle$, $|e\rangle$, $|1\rangle$\} is~\cite{RMPs}
\begin{eqnarray}\label{eq1}
H_{0}=\frac{\hbar}{2}\left[
\begin{array}{ccc}
0                           &\Omega_p(t)                 & 0    \\
\Omega_p(t)^\ast                 &-2\Delta                     &\Omega_s(t)^\ast  \\
0               & \Omega_s(t)                &0
\end{array}
\right].
\end{eqnarray}
Here $\Omega_{p(s)}(t)$~(containing the phase factor) denotes the time-dependent Rabi frequency of the pump~(Stokes) classical field driving the transition $|0\rangle(|1\rangle)\leftrightarrow|e\rangle$. $\Delta=(E_e-E_0)/\hbar-\omega_p-t\dot{\omega}_p=(E_e-E_1)/\hbar-\omega_s-t\dot{\omega}_s$ is the one-photon detuning with $E_j/\hbar$~($j=0,1,e$) and $\omega_{p(s)}$ being frequencies of the bare state $|j\rangle$ and the pump~(Stokes) classical field, respectively.

In such a three-level system, using counterdiabatic driving to speed up STIRAP usually requires an additional coupling between $|0\rangle$ and $|1\rangle$~\cite{XiChen2010} which might be difficult or even impossible to implement in various systems~\cite{YXDu2016,Lin2012OE}. While, such a troublesome problem does not exist in two-level systems~\cite{XiChen2010}, which may provide a solution to this problem of the three-level system. In what follows we shall show three proposals of constructing speeded-up adiabatic USQG by using experimentally feasible counterdiabatic driving in three different cases including time-dependent detuning $\Delta=\Delta(t)$, large detuning $\Delta\gg\Omega_{p,s}(t)$ and one-photon resonance $\Delta=0$, respectively.

\subsection{Time-dependent detuning $\Delta=\Delta(t)$}
\begin{figure*}[htb]\centering
\centering
\includegraphics[width=0.7\linewidth]{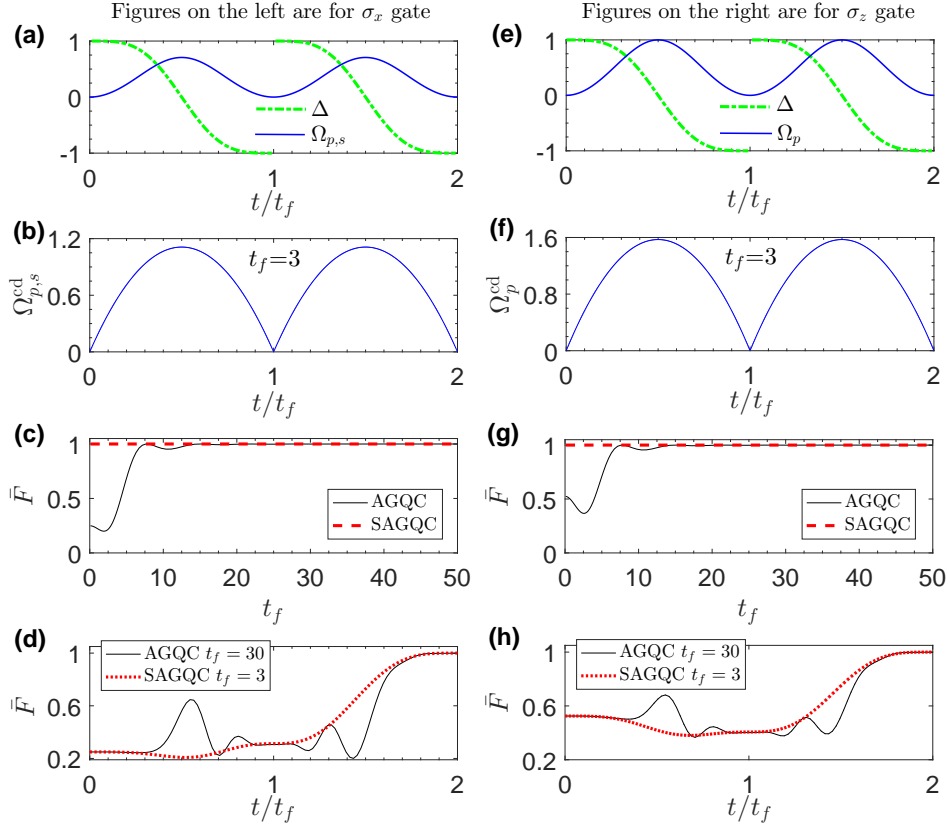}
\caption{Numerical demonstration of $\sigma_x$ gate at time-dependent detuning $\Delta=\Delta(t)$:
(a)~Time-dependent detuning~(green dashed line) and two equal Rabi frequencies~(blue solid line) for the adiabatic $\sigma_x$ gate;
(b)~Counterdiabatic Rabi frequencies $\Omega^{\rm cd}_{p,s}=\Omega^{\rm cd}(t)/\sqrt2$ added into $\Omega_{p,s}(t)$ with $t_f=3$;
(c)~Effect of the value of $t_f$ on the final average fidelity at the time $T=2t_f$ for AGQC~(black solid line) or for SAGQC~(red dashed line);
(d)~Average fidelity trends over time for AGQC~(black solid line) with $t_f=30$ or for SAGQC~(red dotted line) with $t_f=3$.
Numerical demonstration of $\sigma_z$ gate with $\Omega^{\rm cd}_{s}=\Omega_{s}=0$ at time-dependent detuning $\Delta=\Delta(t)$:
(e)~Time-dependent detuning~(green dashed line) and $\Omega_p$~(blue solid line) for the adiabatic $\sigma_z$ gate;
(f)~Counterdiabatic Rabi frequency $\Omega^{\rm cd}_{p}=\Omega^{\rm cd}(t)$ added into $\Omega_{p}(t)$ with $t_f=3$;
(g)~Effect of the value of $t_f$ on the final average fidelity at the time $T=2t_f$ for AGQC~(black solid line) or for SAGQC~(red dashed line);
(h)~Average fidelity trends over time for AGQC~(black solid line) with $t_f=30$ or for SAGQC~(red dotted line) with $t_f=3$.}\label{f1}
\end{figure*}
\subsubsection{Effective two-level system}
The Rabi frequencies of the driving fields could be parameterized as $\Omega_p(t)=e^{-i\varphi}\Omega_\eta(t)\sin\eta$ and $\Omega_s(t)=e^{-i\varphi}\Omega_\eta(t)\cos\eta$ with a same constant phase $\varphi$, $\Omega_\eta(t)$ being real.
In order to construct geometric gates with experimentally feasible counterdiabatic driving, we keep $\eta$ constant to transform $H_{0}$ into a standard two-level form with the basis states \{$|\Phi\rangle\equiv\sin\eta|0\rangle+\cos\eta|1\rangle$, $|e\rangle$\}
\begin{eqnarray}\label{eq2}
H_{\Phi-e}=\frac{\hbar}{2}\left[
\begin{array}{cc}
\Delta(t)                 &\Omega_\eta(t)e^{-i\varphi}\\
\Omega_\eta(t)e^{i\varphi}                 &-\Delta(t)\\
\end{array}
\right].
\end{eqnarray}
The excluded basis state $|d\rangle\equiv\cos\eta|0\rangle-\sin\eta|1\rangle$ is decoupled to $|\Phi\rangle$ and $|e\rangle$, which means that $|d\rangle$ keeps invariant all the time during the system evolution. Further parameterize $\Omega_\eta(t)$ and $\Delta(t)$ by $\Omega_\eta(t)=\Omega_\theta(t)\sin\theta(t)$ and $\Delta(t)=\Omega_\theta(t)\cos\theta(t)$, and then instantaneous eigenstates of $H_{\Phi-e}$ are
\begin{eqnarray}\label{eq3}
|\lambda_+(t)\rangle&=&\cos\frac{\theta(t)}2e^{-i\varphi}|\Phi\rangle+\sin\frac{\theta(t)}2|e\rangle,\nonumber\\
|\lambda_-(t)\rangle&=&-\sin\frac{\theta(t)}2|\Phi\rangle+\cos\frac{\theta(t)}2e^{i\varphi}|e\rangle,
\end{eqnarray}
with corresponding eigenvalues $\lambda_+(t)=\hbar\Omega_\theta(t)/2$ and $\lambda_-(t)=-\hbar\Omega_\theta(t)/2$, respectively.

\subsubsection{Adiabatic USQG}
In the following discussions, we assume the initial time of the quantum-system evolution is $t=0$ and the final time is $t=T$. If we achieve the evolution from $|\Phi\rangle$ at $t=0$ to $e^{i\gamma_\Phi}|\Phi\rangle$ at $t=T$, the unitary operation on the subspace \{$|d\rangle$, $|\Phi\rangle$\} is
\begin{eqnarray}\label{eq4}
U_{d-\Phi}=\left[
\begin{array}{cc}
1              &0\\
0                 &e^{i\gamma_\Phi}\\
\end{array}
\right].
\end{eqnarray}
Thus in the computational space spanned by \{$|0\rangle$, $|1\rangle$\}, we obtain a gate operation
\begin{eqnarray}\label{eq5}
U_{0-1}=\left[
\begin{array}{cc}
\cos^2\eta+e^{i\gamma_\Phi}\sin^2\eta &\cos\eta\sin\eta(e^{i\gamma_\Phi} -1)\\
\cos\eta\sin\eta(e^{i\gamma_\Phi} -1)               &e^{i\gamma_\Phi}\cos^2\eta+\sin^2\eta\\
\end{array}
\right],\nonumber\\
\end{eqnarray}
which can construct a set of USQG by choosing proper $\eta$ and $\gamma_\Phi$. For example, one can set $\eta=\pi/4$ and $\gamma_\Phi=\pi$ to get a $\sigma_x$~(NOT) gate, or $\eta=\pi/2$ and $\gamma_\Phi=\pi$ to get a $\sigma_z$~($\pi$-phase) gate.

In order to achieve the evolution from $|\Phi\rangle$ at $t=0$ to $e^{i\gamma_\Phi}|\Phi\rangle$ at $t=T$, the quantum system may adiabatically and cyclically evolves along a certain eigenstate $|\lambda_+(t)\rangle$ or $|\lambda_-(t)\rangle$ to implement $|\lambda_\pm(0)\rangle=|\Phi\rangle\rightarrow|\lambda_\pm(T)\rangle=e^{i(a_\pm+\gamma_\pm)}|\Phi\rangle$, with $a_\pm=-\int_0^T\lambda_\pm(t)dt$ being the dynamical phase and $\gamma_\pm=i\int_0^T\langle\lambda_\pm(t)|\partial_t\lambda_\pm(t)\rangle dt$ being the geometric phase. However, the phase factor containing the non-zero dynamical phase $a_\pm$ is not the desired result at all. To erase the accumulated dynamical phase and obtain an all-geometric phase, we suggest to adopt a double-mode adiabatic-evolution path ``$|\lambda_+(t)\rangle$ and $|\lambda_-(t)\rangle$" instead of ``$|\lambda_+(t)\rangle$ or $|\lambda_-(t)\rangle$", as the following two steps
\begin{eqnarray}\label{eq6}
&&{\rm step~1}:~|\lambda_\pm(0)\rangle=|\Phi\rangle\rightarrow|\lambda_\pm(T/2)\rangle=|e\rangle,\nonumber\\
&&{\rm step~2}:~|\lambda_\mp(T/2)\rangle=|e\rangle\rightarrow|\lambda_\mp(T)\rangle=e^{\pm i\Gamma_\Phi}|\Phi\rangle,
\end{eqnarray}
where $\Gamma_\Phi=\pi+\varphi_1-\varphi_2$ is a purely geometric phase with $\varphi_1$ and $\varphi_2$ being the constant phase of the classical-field driving during the step~1 and step~2, respectively. Because of $\lambda_+(t)=-\lambda_-(t)$, the accumulated dynamical phase during the total quantum-system evolution is zero.

\subsubsection{Speeded-up adiabatic USQG}
The procedure above is based on adiabatic evolution, slow as we all know. According to the theory of counterdiabatic driving, the procedure above of constructing USQG can be speeded up by adding a counterdiabatic Hamiltonian $H_{\rm cd}$ into $H_{\Phi-e}$. The form of $H_{\rm cd}$ is~\cite{XiChen2010}
\begin{eqnarray}\label{eq7}
H_{\rm cd}=\frac{\hbar}{2}\left[
\begin{array}{cc}
0                 &-i\Omega^{\rm cd}(t)e^{-i\varphi}\\
i\Omega^{\rm cd}(t)e^{i\varphi}                 &0\\
\end{array}
\right],
\end{eqnarray}
where $\Omega^{\rm cd}(t)\equiv[\dot{\Omega}_\eta(t)\Delta(t)-\Omega_\eta(t)\dot{\Delta}(t)]/\Omega_\theta(t)^2$.
Then we obtain a modified Hamiltonian $H_S=H_{\Phi-e}+H_{\rm cd}$ that could govern the quantum system exactly~(not slowly) evolving along a certain eigenstate in Eq.~(\ref{eq3}). Besides, $H_{\rm cd}$ is experimentally feasible because it does not introduce the coupling between $|0\rangle$ and $|1\rangle$ but just modify the classical-field pulse shapes and phases, as
\begin{eqnarray}\label{eq8}
\Omega_p(t)&\rightarrow&\Omega_p^{\rm m}(t)=[\Omega_\eta(t)-i\Omega^{\rm cd}(t)]\sin\eta e^{-i\varphi},\nonumber\\
\Omega_s(t)&\rightarrow&\Omega_s^{\rm m}(t)=[\Omega_\eta(t)-i\Omega^{\rm cd}(t)]\cos\eta e^{-i\varphi}.
\end{eqnarray}
In a word, according to the procedure in Eq.~(\ref{eq6}), one can construct speeded-up adiabatic all-geometric USQG in Eq.~(\ref{eq5}) by using the modified Rabi frequencies in Eq.~(\ref{eq8}).

\subsubsection{Numerical demonstration}
For demonstrating the effectiveness of constructing speeded-up USQG, we set $\theta(0)=0\rightarrow\theta(T_-/2)=\pi\rightarrow\theta(T_+/2)=0\rightarrow\theta(T)=\pi$
to meet the procedure in Eq.~(\ref{eq6}), $T_-/2$ and $T_+/2$~(mathematically, $T_-/2=T_+/2=T/2$) denoting the ending time of the step~1 and the beginning time of the step~2, respectively. The form of $\theta(t)$ can be expressed by partitioned polynomial ansatz. All needed parameters are listed in table~\ref{T1}, for which we adopt the dimensionless natural unit~($\hbar=1$), and also we choose $\Omega_\theta(t)=1$, for simplicity. In Fig.~\ref{f1}, we show the results of simulating $\sigma_x$ gate~[(a)-(d)] and $\sigma_z$ gate~[(e)-(h)], for which the average fidelity of the USQG in Eq.~(\ref{eq5}) is defined as
\begin{eqnarray}\label{eq9}
\bar{F}(t)=\frac1{4\pi^2}\int_0^{2\pi}\int_0^{2\pi}|\langle\Psi_U|\Psi(t)\rangle|^2d\alpha_1d\alpha_2,
\end{eqnarray}
where $|\Psi(t)\rangle$ is the state of the three-level system governed by the Sch\"{o}dinger equation based on the Hamiltonian in Eq.~(\ref{eq1}) with an arbitrary initial state $|\Psi(0)\rangle=\sin\alpha_1|0\rangle+\cos\alpha_1e^{i\alpha_2}|1\rangle$, and $|\Psi_U\rangle=U_{0-1}|\Psi(0)\rangle$ is the target state after executing the USQG $U_{0-1}$ on $|\Psi(0)\rangle$.
\begin{table}
\centering
\caption{\label{T1}Parameters of constructing USQG with $\Delta=\Delta(t)$. Preset parameters: $\hbar=1$ and $\Omega_\theta(t)=1$.}
\footnotesize
\begin{tabular}{@{}ll}
\hline\hline
Parameter&Expression\\
\hline
$T$&$2t_f$\\

$\theta$&$\left\{\begin{array}{ll}
\frac{3\pi t^2}{{t_f}^2}-\frac{2\pi t^3}{{t_f}^3},&0\leq t<t_f\\
\frac{3\pi (t-t_f)^2}{{t_f}^2}-\frac{2\pi (t-t_f)^3}{{t_f}^3},&t_f\leq t<2t_f
\end{array}\right.$\\

$\varphi$&$0$\\

$\eta$&$\left\{\begin{array}{ll}
\pi/4,&\sigma_x\\
\pi/2,&\sigma_z
\end{array}\right.$\\
\hline\hline
\end{tabular}
\end{table}

 About the amplitudes, $\Omega_p(t)/\Omega_s(t)=\tan\eta$ makes the two Rabi frequencies $\Omega_{p,s}(t)$, as well as $\Omega_{p,s}^{\rm cd}(t)$~{with a $\pi$-phase difference from $\Omega_{p,s}(t)$} added into $\Omega_{p,s}(t)$, keep a constant ratio that is determined by the value of $\eta$, as shown in Figs.~\ref{f1}(a), (b), (e) and (f). The maximum amplitudes $\max\{\Omega_{p,s}(t)\}$ determined by $\Omega_\theta$ and $\eta$ are fixed with an arbitrary $t_f$, while $\max\{\Omega_{p,s}^{\rm cd}(t)\}$ determined by $\dot{\theta}(t)$ would be inversely proportional to $t_f$. Therefore, a suitable $t_f$~(i.e., a half of operation time) that determines the intensity of the classical fields~(costs of energy) should be picked in practice, although $t_f$ could be indefinitely short in principle~[see Figs.~\ref{f1}(c) and (g)] to guarantee a unity fidelity. A unity fidelity for AGQC requires the value of $t_f$ large enough to satisfy the adiabatic condition, as shown in Figs.~\ref{f1}(c) and (g). For $\max\{\Omega_{p,s}^{\rm cd}(t)\}$ comparable to $\max\{\Omega_{p,s}(t)\}$, we pick $t_f=3$ to implement SAGQC. To implement high-fidelity AGQC similar to SAGQC, $t_f$ needs to be raised by an order of magnitude, as shown in Figs.~\ref{f1}(d) and (h). The discussion above shows the effectiveness of constructing speeded-up adiabatic USQG with time-dependent detuning.

\subsection{Large detuning $\Delta\gg\Omega_{p,s}(t)$}

\subsubsection{Effective two-level system}
At large intermediate-level detuning $\Delta\gg\Omega_{p,s}(t)$, the state $|e\rangle$ is populated little, and it could be adiabatically eliminated~\cite{Brion2007,James2007}. Then the following effective two-level Hamiltonian can be obtained with the basis states \{$|0\rangle$, $|1\rangle$\}~\cite{Vitanov1997}
\begin{eqnarray}\label{eq10}
H_{\rm eff}=\frac{\hbar}{2}\left[
\begin{array}{cc}
\Delta_{\rm eff}(t)                 &\Omega_{\rm eff}(t)e^{-i\varphi}\\
\Omega_{\rm eff}(t)e^{i\varphi}                 &-\Delta_{\rm eff}(t)\\
\end{array}
\right],
\end{eqnarray}
with the effective detuning and Rabi frequency being
\begin{eqnarray}\label{eq11}
\Delta_{\rm eff}(t)=\frac{\Omega_p(t)^2-\Omega_s(t)^2}{4\Delta} {\rm~ and~}
\Omega_{\rm eff}(t)=\frac{\Omega_p(t)\Omega^\ast_s(t)}{2\Delta},\nonumber
\end{eqnarray}
respectively.
In order to introduce a phase factor into $H_{\rm eff}$, we set $\Omega_{p,s}(t)$ real and $e^{i\varphi}$ accompanying with $\Omega_{s}(t)$. We learn that $H_{\rm eff}$ has the same form as $H_{\Phi-e}$ in Eq.~(\ref{eq2}) but is with different basis states from $H_{\Phi-e}$. Therefore, the eigenstates of $H_{\rm eff}$ have the same form as those of $H_{\Phi-e}$ but with basis states \{$|0\rangle$, $|1\rangle$\}
\begin{eqnarray}\label{eq12}
|\Lambda_+(t)\rangle&=&\cos\frac{\Theta(t)}2e^{-i\varphi}|0\rangle+\sin\frac{\Theta(t)}2|1\rangle,\nonumber\\
|\Lambda_-(t)\rangle&=&-\sin\frac{\Theta(t)}2|0\rangle+\cos\frac{\Theta(t)}2e^{i\varphi}|1\rangle,
\end{eqnarray}
with corresponding eigenvalues $\Lambda_+(t)=\hbar\Omega_\Theta(t)/2$ and $\Lambda_-(t)=-\hbar\Omega_\Theta(t)/2$, respectively. We define the mixing angle $\Theta=\arctan[\Omega_{\rm eff}(t)/\Delta_{\rm eff}(t)]$ and mixing amplitude $\Omega_\Theta(t)=\sqrt{\Omega_{\rm eff}(t)^2+\Delta_{\rm eff}(t)^2}$.

The effective Hamiltonian $H_{\rm eff}$ has been researched for speeding up population transfer of STIRAP by counterdiabatic driving in Refs.~\cite{YCLi2016,YXDu2017,Mortensen2018}, and the corresponding experimental realization has also been done in a cold-atom system~\cite{YXDu2016}. With the aid of this effective two-level Hamiltonian, now we construct speeded-up adiabatic USQG.

\subsubsection{Adiabatic USQG}
In order to construct adiabatic USQG and erase the accumulated dynamical phase, we could follow the double-mode adiabatic-evolution path ``$|\Lambda_+(t)\rangle$ and $|\Lambda_-(t)\rangle$" just similar to the case of the time-dependent detuning, as the following two steps
\begin{eqnarray}\label{eq13}
&&{\rm step~1}:~|\Lambda_\pm(0)\rangle\rightarrow|\Lambda_\pm(T_-/2)\rangle,\nonumber\\
&&{\rm step~2}:~|\Lambda_\mp(T_+/2)\rangle\rightarrow|\Lambda_\mp(T)\rangle=e^{\pm i\widetilde{\Gamma}}|\Lambda_\pm(0)\rangle,\nonumber\\
\end{eqnarray}
with $\widetilde{\Gamma}$ being a purely geometric phase. Obviously, $|\Lambda_\pm(T_-/2)\rangle$ and $|\Lambda_\mp(T_+/2)\rangle$ are supposed to describe a same quantum state~(ignoring the global phase), which can be enabled by choosing an appropriate form of $\Theta(t)$.
The unitary evolution operator achieving $\widetilde{U}|\Lambda_\pm(0)\rangle=e^{\pm i\widetilde{\Gamma}}|\Lambda_\pm(0)\rangle$ in the subspace \{$|\Lambda_+(0)\rangle$, $|\Lambda_-(0)\rangle$\} is
\begin{eqnarray}\label{eq14}
\widetilde{U}=\left[
\begin{array}{cc}
e^{i\widetilde{\Gamma}}              &0\\
0                 &e^{-i\widetilde{\Gamma}}\\
\end{array}
\right].
\end{eqnarray}
Thus in the computational space spanned by \{$|0\rangle$, $|1\rangle$\}, we obtain a gate operation~\cite{SLZhu}
\begin{eqnarray}\label{eq15}
\widetilde{U}_{0-1}=\left[
\begin{array}{cc}
\cos\widetilde{\Gamma}+i\cos\widetilde{\Theta}\sin\widetilde{\Gamma}               &i\sin\widetilde{\Theta}\sin\widetilde{\Gamma}\\
i\sin\widetilde{\Theta}\sin\widetilde{\Gamma}              &\cos\widetilde{\Gamma}-i\cos\widetilde{\Theta}\sin\widetilde{\Gamma} \\
\end{array}
\right],\nonumber\\
\end{eqnarray}
which can be used to construct the gate $\widetilde{U}_1=e^{i\widetilde{\Gamma}\sigma_z}$ by choosing the initial value $\widetilde{\Theta}\equiv{\Theta}(0)=0$ and $\widetilde{U}_2=e^{i\widetilde{\Gamma}\sigma_x}$ by $\widetilde{\Theta}=\pi/2$~\cite{ZTLiang2016,YXDu2017}. Due to the noncommutability between $\widetilde{U}_1$ and $\widetilde{U}_2$, one can get any single-qubit operation by combining $\widetilde{U}_1$ and $\widetilde{U}_2$.

It is worth noting that engineering directly adiabatic Rabi frequencies $\Omega_{p,s}(t)$ to perfectly perform the procedure in Eq.~(\ref{eq13}) and thus implement $\widetilde{U}_{0-1}$ in Eq.~(\ref{eq15}) is pretty difficult, and also may cause the difficulty of implementing speeded-up adiabatic USQG. Therefore, it shall be better to engineer $\Omega_{\rm eff}(t)$ and $\Delta_{\rm eff}(t)$ firstly, and then calculate inversely the adiabatic Rabi frequencies
\begin{eqnarray}\label{eq16}
\Omega_{p}(t)&=&\sqrt{2\Delta[\sqrt{\Delta_{\rm eff}(t)^2+\Omega_{\rm eff}(t)^2}+\Delta_{\rm eff}(t)]},\nonumber\\
\Omega_{s}(t)&=&\sqrt{2\Delta[\sqrt{\Delta_{\rm eff}(t)^2+\Omega_{\rm eff}(t)^2}-\Delta_{\rm eff}(t)]}.
\end{eqnarray}
Since Eq.~(\ref{eq16}) involves many square and square-root calculations that may lead to multiple solutions not satisfying Eq.~(\ref{eq11}) and then not implementing $\widetilde{U}_{0-1}$, $\Omega_{\rm eff}(t)$ and $\Delta_{\rm eff}(t)$ should be designed very carefully.
\subsubsection{Speeded-up adiabatic USQG}
In a two-level entity system~(NV center), the implementation of the speeded-up adiabatic $\widetilde{U}_{0-1}$ by counterdiabatic driving has been discussed in  Ref.~\cite{ZTLiang2016}, and the corresponding experimental realization has also been published not long ago~\cite{FK2018}. However, in a three-level entity system with the Hamiltonian Eq.~(\ref{eq1}), the implementation of the speeded-up adiabatic $\widetilde{U}_{0-1}$ by counterdiabatic driving is of much more difficulties than those in a two-level entity system, because the realistically used Rabi frequencies are inversely deduced by the effective detuning and Rabi frequency~[see Eq.~(\ref{eq16})] instead of being engineered directly. The large-detuning-case USQG $\widetilde{U}_{0-1}$ in Eq.~(\ref{eq15}) has been reported recently with the methods of STIRAP and counterdiabatic-driving STA in Ref.~\cite{YXDu2017}. The study~\cite{YXDu2017} mainly aims at geometric atom interferometry, so just ``population dynamics"~(population transfer between two lower-energy states) is discussed in details, which may be inadequate for the construction of quantum gates.

For the effective Hamiltonian $H_{\rm eff}$ in Eq.~(\ref{eq10}), the modified effective Rabi frequency replacing ${\Omega}_{\rm eff}(t)$ by counterdiabatic driving  becomes ${\Omega}'_{\rm eff}(t)\equiv{\Omega}_{\rm eff}(t)-i\Omega_{\rm eff}^{\rm cd}(t)$ with $\Omega_{\rm eff}^{\rm cd}(t)\equiv\dot{\Theta}(t)$.
With the results of the inverse calculations in Eq.~(\ref{eq16}) with ${\Omega'}_{\rm eff}(t)$ replacing ${\Omega}_{\rm eff}(t)$, it is indeterminate to conversely satisfy
${\Omega'}_{\rm eff}(t)={{\Omega}_p(t){\Omega}_s^\ast(t)}/{2\Delta}$ that is definitely necessary because the effective two-level Hamiltonian must be from the Hamiltonian of the three-level system. In this proposal, therefore, we would not choose easily this way to speed up the adiabatic $\widetilde{U}_{0-1}$.

For the implementation of the speeded-up adiabatic $\widetilde{U}_{0-1}$, as a matter of fact, the counterdiabatic Hamiltonian $H_{\rm eff}^{\rm cd}$ alone instead of $H_{\rm eff}^{\rm cd}+H_{\rm eff}$ is enough according to the theory of TQD~\cite{Berry2009,XiChen2010}. Besides, the global phase factor $i$ could be omitted. Thus the modified effective two-level Hamiltonian becomes
\begin{eqnarray}\label{eq17}
{H}_{\rm eff}^{\rm m}=\frac{\hbar}{2}\left[
\begin{array}{cc}
0                 &\Omega_{\rm eff}^{\rm cd}(t)e^{-i\varphi}\\
\Omega_{\rm eff}^{\rm cd}(t)e^{i\varphi}                 &0\\
\end{array}
\right].
\end{eqnarray}
Accordingly, the realistically-used modified Rabi frequencies replacing ${\Omega}_{p,s}(t)$ in the three-level system are
\begin{eqnarray}\label{eq18}
{\Omega}_p^{\rm cd}(t)&=&\sqrt{2\Delta[\sqrt{\Delta_{\rm eff}(t)^2+\Omega_{\rm eff}^{\rm cd}(t)^2}+\Delta_{\rm eff}(t)]},\nonumber\\
{\Omega}_s^{\rm cd}(t)&=&\sqrt{2\Delta[\sqrt{\Delta_{\rm eff}(t)^2+\Omega_{\rm eff}^{\rm cd}(t)^2}-\Delta_{\rm eff}(t)]},
\end{eqnarray}
which are relatively easy to conversely satisfy $\Omega_{\rm eff}^{\rm cd}(t)={{\Omega}_p^{\rm cd}(t){\Omega}_s^{\rm cd\ast}(t)}/{2\Delta}$ by designing $\Theta(t)$. On the other hand, the fact that ${H}_{\rm eff}^{\rm m}$ in Eq.~(\ref{eq17}) must be from the Hamiltonian of the three-level system also means the modified effective detuning $[{{\Omega}_p^{\rm cd}(t)^2-{\Omega}_s^{\rm cd}(t)^2}]/{4\Delta}=0$, i.e., ${\Omega}_p^{\rm cd}(t)={\Omega}_s^{\rm cd}(t)$. From Eq.~(\ref{eq18}), we learn that ${\Omega}_p^{\rm cd}(t)={\Omega}_s^{\rm cd}(t)$ will be possible when $\Omega_{\rm eff}^{\rm cd}(t)\gg\Delta_{\rm eff}(t)$ is met, which can be achieved by enlarging $\dot{\Theta}(t)$. It just fits with our purpose of shortening time very well. But then $\Omega_{p,s}^{\rm cd}(t)\approx\sqrt{2\Delta\Omega_{\rm eff}^{\rm cd}(t)}$ under the condition $\Omega_{\rm eff}^{\rm cd}(t)\gg\Delta_{\rm eff}(t)$ must obey the premise of large detuning $\Omega_{p,s}^{\rm cd}(t)\ll\Delta$. In a word, according to the procedure in Eq.~(\ref{eq13}), one can construct speeded-up adiabatic USQG in Eq.~(\ref{eq15}) by using the modified Rabi frequencies in Eq.~(\ref{eq18}) under the conditions $\Delta\gg\Omega_{p,s}^{\rm cd}(t)$ and $\Omega_{\rm eff}^{\rm cd}(t)\gg\Delta_{\rm eff}(t)$.

\subsubsection{Numerical demonstration}
\begin{figure*}[htb]\centering
\centering
\includegraphics[width=0.7\linewidth]{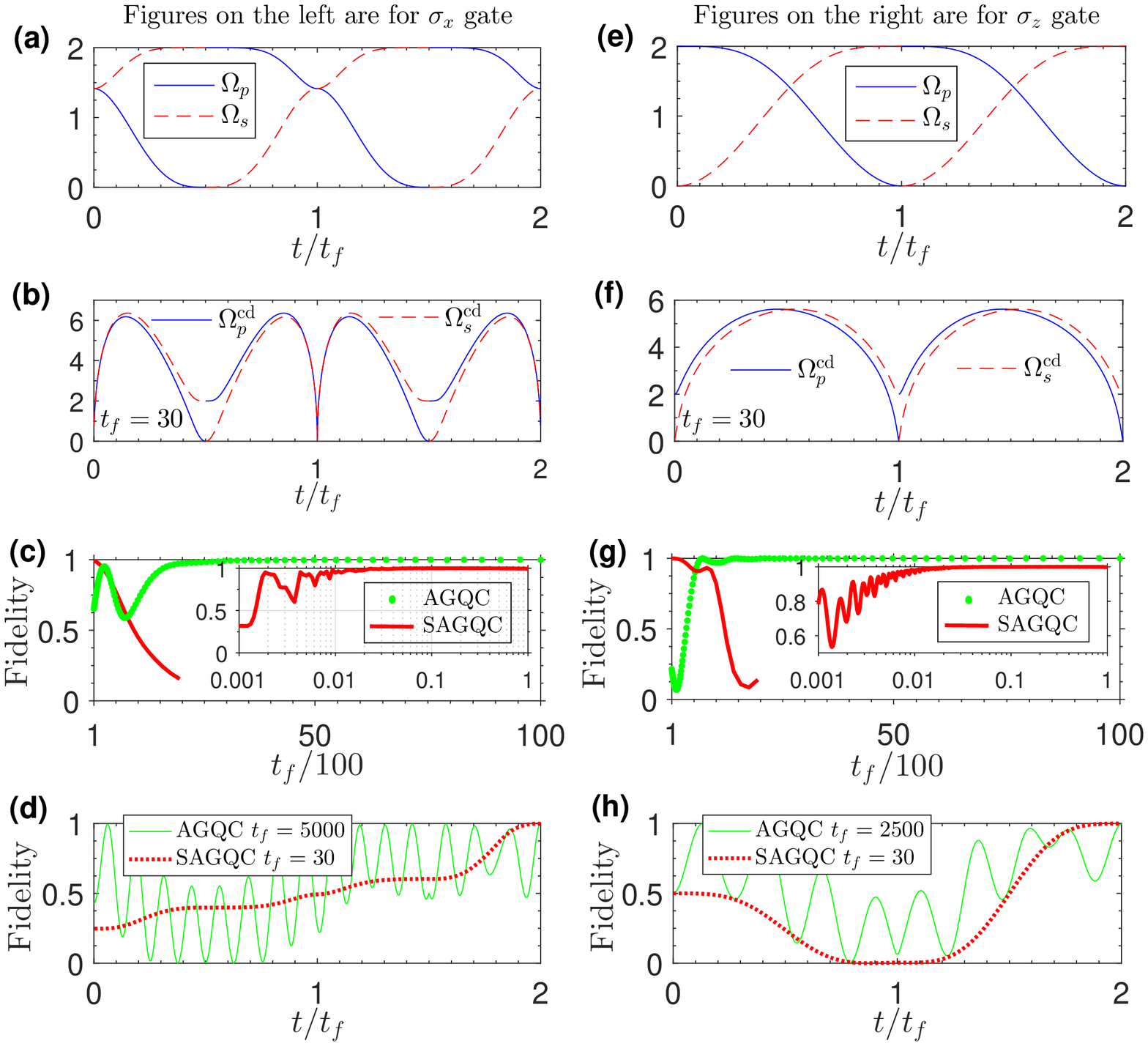}
\caption{Numerical demonstration of $\sigma_x$ gate at large detuning $\Delta\gg\Omega_{p,s}(t)$:
(a)~Time-dependent Rabi frequencies for the adiabatic $\sigma_x$ gate;
(b)~Counterdiabatic Rabi frequencies with $t_f=30$;
(c)~Effect of the value of $t_f$ on the final fidelity at the time $T=2t_f$ for AGQC~(green dotted line) or for SAGQC~(red solid line);
(d)~Fidelity trends over time for AGQC~(green solid line) with $t_f=5000$ or for SAGQC~(red dotted line) with $t_f=30$.
Numerical demonstration of $\sigma_z$ gate at large detuning $\Delta\gg\Omega_{p,s}(t)$:
(e)~Time-dependent Rabi frequencies for the adiabatic $\sigma_z$ gate;
(f)~Counterdiabatic Rabi frequencies with $t_f=30$;
(g)~Effect of the value of $t_f$ on the final fidelity at the time $T=2t_f$ for AGQC~(green dotted line) or for SAGQC~(red solid line);
(h)~Fidelity trends over time for AGQC~(green solid line) with $t_f=2500$ or for SAGQC~(red dotted line) with $t_f=30$.}\label{f2}
\end{figure*}

For demonstrating the effectiveness of constructing speeded-up USQG in Eq.~(\ref{eq15}), we set $\Theta(0)=-\pi/2\rightarrow\Theta(T_-/2)=\pi/2\rightarrow\Theta(T_+/2)=-\pi/2\rightarrow\Theta(T)=\pi/2$ for $\sigma_x$ gate and $\Theta(0)=0\rightarrow\Theta(T_-/2)=\pi\rightarrow\Theta(T_+/2)=0\rightarrow\Theta(T)=\pi$ for $\sigma_z$ gate, respectively. All needed parameters are listed in table~\ref{T2} for $\sigma_x$ gate and table~\ref{T3} for $\sigma_z$ gate, respectively. We still adopt the dimensionless natural unit~($\hbar=1$), and we choose $\Omega_\Theta(t)=0.01$ and $\Delta=50$ to obey the condition of large detuning. In Fig.~\ref{f2}, we show the results of simulating $\sigma_x$ gate~[(a)-(d)] and $\sigma_z$ gate~[(e)-(h)]. Because it needs more run time for simulating the case of the large detuning, the fidelity of the USQG in Eq.~(\ref{eq15}) is defined as
$F(t)=|\langle\widetilde{\Psi}_U|\Psi(t)\rangle|^2$, where $|\Psi(t)\rangle$ is the state of the three-level system governed by the Sch\"{o}dinger equation based on the Hamiltonian~(\ref{eq1}). The initial state is $|\Psi(0)\rangle=\sin\alpha_1|0\rangle+\cos\alpha_1e^{i\alpha_2}|1\rangle$ with $\alpha_1=\pi/8$ and $\alpha_2=3\pi/8$, without loss of generality. $|\widetilde{\Psi}_U\rangle=\widetilde{U}_{0-1}|\Psi(0)\rangle$ is the target state after executing the USQG $\widetilde{U}_{0-1}$ on $|\Psi(0)\rangle$.
\begin{table}
\centering
\caption{\label{T2}Parameters of constructing $\sigma_x$ with $\Delta\gg\Omega_{p,s}(t)$. Preset parameters: $\hbar=1$, $\Omega_\Theta(t)=0.01$, $\Delta=50$, $a_0=\pi/2$, $a_1=0$, $a_2=-15\pi/{t_f}^2$, $a_3=50\pi/{t_f}^3$, $a_4=-60\pi/{t_f}^4$ and $a_5=24\pi/{t_f}^5$.}
\footnotesize
\begin{tabular}{@{}ll}
\hline\hline
Parameter&Expression\\
\hline
$T$&$2t_f$\\

$\Theta_x$&$\left\{\begin{array}{ll}
\sum_{k=0}^5a_kt^k,&0\leq t<t_f\\
\sum_{k=0}^5a_k(t-t_f)^k,&t_f\leq t<2t_f
\end{array}\right.$\\

$\varphi$&$\left\{\begin{array}{ll}
0,&0\leq t<t_f/2~\&~\frac{3t_f}2\leq t<2t_f\\
\pi/2,&\frac{t_f}2\leq t<\frac{3t_f}2
\end{array}\right.$\\

$\Delta_{\rm eff}$&$\left\{\begin{array}{ll}
\Omega_\Theta\cos(\pi+\Theta_x),&0\leq t<\frac{t_f}2~\&~t_f\leq t<\frac{3t_f}2\\
\Omega_\Theta\cos\Theta_x,&\frac{t_f}2\leq t<t_f~\&~\frac{3t_f}2\leq t<2t_f
\end{array}\right.$\\

$\Omega_{\rm eff}$&$\left\{\begin{array}{ll}
\Omega_\Theta\sin\Theta_x,&0\leq t<\frac{t_f}2~\&~t_f\leq t<\frac{3t_f}2\\
\Omega_\Theta\sin(-\Theta_x),&\frac{t_f}2\leq t<t_f~\&~\frac{3t_f}2\leq t<2t_f
\end{array}\right.$\\
\hline\hline
\end{tabular}
\end{table}

\begin{table}
\centering
\caption{\label{T3}Parameters of constructing $\sigma_x$ with $\Delta\gg\Omega_{p,s}(t)$. Preset parameters: $\hbar=1$, $\Omega_\Theta(t)=0.01$ and $\Delta=50$.}
\footnotesize
\begin{tabular}{@{}ll}
\hline\hline
Parameter&Expression\\
\hline
$T$&$2t_f$\\

$\Theta_z$&$\left\{\begin{array}{ll}
\frac{3\pi t^2}{{t_f}^2}-\frac{2\pi t^3}{{t_f}^3},&0\leq t<t_f\\
\frac{3\pi (t-t_f)^2}{{t_f}^2}-\frac{2\pi (t-t_f)^3}{{t_f}^3},&t_f\leq t<2t_f
\end{array}\right.$\\

$\varphi$&$\left\{\begin{array}{ll}
0,&0\leq t<t_f\\
\pi/2,&t_f\leq t<2t_f
\end{array}\right.$\\

$\Delta_{\rm eff}$&$\Omega_\Theta\cos\Theta_z$\\

$\Omega_{\rm eff}$&$\Omega_\Theta\sin\Theta_z$\\
\hline\hline
\end{tabular}
\end{table}

For the adiabatic $\sigma_x$ gate~[Fig.~\ref{f2}(a)] and $\sigma_z$ gate~[Fig.~\ref{f2}(e)], the maximum amplitudes $\max\{\Omega_{p,s}(t)\}$ are comparable~(around two times) with those in the case of time-dependent detuning, which means that we can contrast the operation time of the two cases. In Figs.~\ref{f2}(c) and (g), obviously, the order of magnitude of the operation time is at least $10^3$ for the adiabatic USQG~(green dotted lines) or at least $10^1$ for the speeded-up adiabatic USQG~(red solid lines) to keep a high fidelity, which illustrates that the speeded-up adiabatic USQG could reduce the operation time by two orders of magnitude. Besides, for the adiabatic USQG, compared with the case of time-dependent detuning, the operation time of the case of large detuning increases by two orders of magnitude. For the speeded-up adiabatic USQG at large detuning, the effect of the value of $t_f$ on the final fidelity~[red solid lines in Figs.~\ref{f2}(c) and (g)] is much different from that at time-dependent detuning. The final fidelity increases up to unity with the increase of $t_f$ while then decreases when $t_f$ is over a certain value~(about $10^2$), which is accompanying with the fact that with the increase of $t_f$, $\Omega_{p,s}^{\rm cd}(t)$ decrease to satisfy the condition $\Delta\gg\Omega_{p,s}^{\rm cd}(t)$ better and better but then will spoil the condition $\Omega_{\rm eff}^{\rm cd}(t)\gg\Delta_{\rm eff}(t)$ when $t_f$ is over a certain value. Fortunately, there is a wide range of the values of $t_f$ keeping a near-unity fidelity. We pick $t_f=30$ to plot the counterdiabatic Rabi frequencies $\Omega_{p,s}^{\rm cd}(t)$ in Figs.~\ref{f2}(b) and (f) for $\sigma_x$ gate and $\sigma_z$ gate, respectively. The condition ${\Omega}_p^{\rm cd}(t)\simeq{\Omega}_s^{\rm cd}(t)$ is roughly satisfied. The maximum amplitudes $\max\{\Omega_{p,s}^{\rm cd}(t)\}$ are of the same order of magnitude as $\max\{\Omega_{p,s}(t)\}$, which guarantees the feasibility of the proposal. More than that in the case of time-dependent detuning,
to implement high-fidelity AGQC similar to SAGQC, $t_f$ needs to be raised by two orders of magnitude, as shown in Figs.~\ref{f2}(d) and (h). The discussion above shows the effectiveness of constructing speeded-up adiabatic USQG at large detuning.

\subsection{One-photon resonance $\Delta=0$}
As for the case of the one-photon resonance $\Delta=0$ in the three-level system, it is unnecessary to implement SAGQC by accelerating AGQC with $\Delta=0$. As we mention in the previous large-detuning subsection, the counterdiabatic Hamiltonian $H^{\rm cd}$ alone instead of $H^{\rm cd}+H_{0}$ is enough to speed up the adiabatic evolution of the three-level system with the Hamiltonian $H_{0}$. Back to the the time-dependent-detuning subsection, hence
\begin{eqnarray}\label{eq19}
H^{\rm cd}=\frac{\hbar}{2}\left[
\begin{array}{cc}
0                 &\Omega^{\rm cd}(t)e^{-i\varphi}\\
\Omega^{\rm cd}(t)e^{i\varphi}                 &0\\
\end{array}
\right]
\end{eqnarray}
is enough to implement speeded-up adiabatic USQG Eq.~(\ref{eq5}) by taking the place of ($H_{\Phi-e}+H_{\rm cd}$),
where $H_{\Phi-e}$ and $H_{\rm cd}$ are in Eqs.~(\ref{eq2}) and (\ref{eq7}), respectively. In the three-level system, the Hamiltonian performing speeded-up adiabatic USQG Eq.~(\ref{eq5}) becomes
\begin{eqnarray}\label{eq20}
H_{0}^{\rm cd}=\frac{\hbar\Omega^{\rm cd}(t)}{2}\left[
\begin{array}{ccc}
0                           &\sin\eta e^{-i\varphi}                 & 0    \\
\sin\eta e^{i\varphi}                 &0                     &\cos\eta e^{i\varphi}  \\
0               &\cos\eta e^{-i\varphi}                &0
\end{array}
\right],\nonumber\\
\end{eqnarray}
which is just the Hamiltonian of a standard one-photon-resonance three-level system. $\Omega^{\rm cd}(t)$ has been defined below Eq.~(\ref{eq7}).

Alternatively, the speeded-up adiabatic USQG Eq.~(\ref{eq5}) in the case of one-photon resonance can be implemented by inverse engineering. We know that the double paths in Eq.~(\ref{eq3}) do not satisfy the Schr\"{o}dinger equation based on $H_{\Phi-e}$ Eq.~(\ref{eq2})  without the adiabatic evolution, so the procedure Eq.~(\ref{eq6}) can not be followed. In order to implement speeded-up adiabatic USQG Eq.~(\ref{eq5}), one can inversely engineer a Hamiltonian $H_{\rm en}$ replacing $H_{\Phi-e}$ to make the double paths satisfy $i\hbar\partial_t|\lambda_\pm(t)\rangle=H_{\rm en}|\lambda_\pm(t)\rangle$. The form of $H_{\rm en}$ can be chosen easily
\begin{eqnarray}\label{eq020}
H_{\rm en}=i[|\partial_t\lambda_+(t)\rangle\langle\lambda_+(t)|+|\partial_t\lambda_-(t)\rangle\langle\lambda_-(t)|].
\end{eqnarray}
Substituting Eq.~(\ref{eq3}) into Eq.~(\ref{eq020}), we find $H_{\rm en}$ happens to be $H_{\rm cd}$ in Eq.~(\ref{eq7}) that is Eq.~(\ref{eq20}) back to the three-level system~(omitting the global phase factor $i$). Therefore, the one-photon-resonance Hamiltonian $H_{0}^{\rm cd}$ in Eq.~(\ref{eq20}) could govern exactly the system to follow the procedure Eq.~(\ref{eq6}) so as to implement speeded-up adiabatic USQG Eq.~(\ref{eq5}).

In a word, the proposal of speeded-up adiabatic USQG in a one-photon-resonance three-level system is proposed. Numerical demonstration for this case is exactly the same as that for the case of the time-dependent detuning~(see Fig.~\ref{f1}) with the parameters in table~\ref{T1}.

\section{Nontrivial two-qubit gates~(NTQG)}\label{S3}
\begin{figure}
\includegraphics[width=\linewidth]{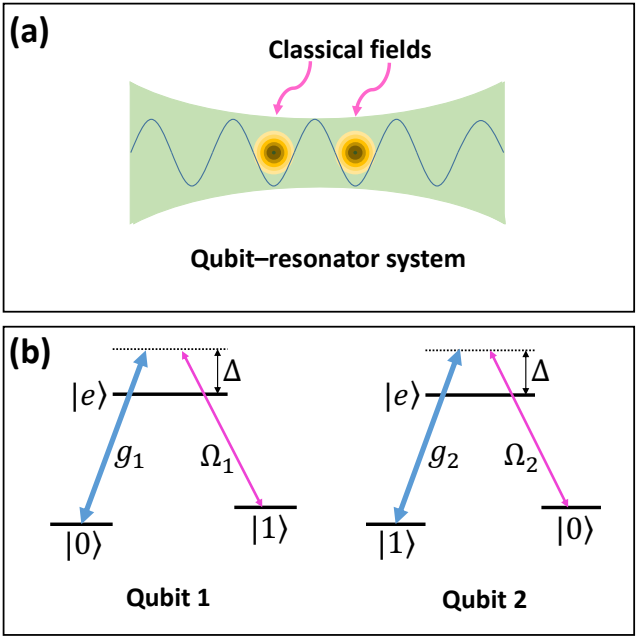}
\caption{(a) The schematic sketch of the qubit-resonator system. Two external classical fields are imposed on two qubits coupled to a single-mode resonator, respectively. (b) Level configurations of the two qubits and the related transitions. The two qubits both possess one high-energy and two lower-energy levels. Binary digits are encoded on the two stable lower-energy levels of the two qubits, antisymmetrically.
Transitions $|0\rangle\leftrightarrow|e\rangle$ of Qubit~1 and $|1\rangle\leftrightarrow|e\rangle$ of Qubit~2 are coupled to the mode of the resonator with coupling constants $g_1$ and $g_2$, respectively. Transitions $|1\rangle\leftrightarrow|e\rangle$ of Qubit~1 and $|0\rangle\leftrightarrow|e\rangle$ of Qubit~2 are driven by the external classical fields with time-dependent Rabi frequencies $\Omega_1(t)$ and $\Omega_2(t)$, respectively. All the four transitions are with the same intermediate-level detuning $\Delta$.}\label{f3}
\end{figure}
As for the implementation of speeded-up adiabatic NTQG, we still use the method of counterdiabatic driving in three-level systems including three cases, i.e., time-dependent detuning, large detuning and one-photon resonance. Consider a modeled qubit-resonator system, and the schematic sketch of the model and level configurations of the two qubits involving related transitions are shown in Figs.~\ref{f3}(a) and (b), respectively. In a concrete physical implementation, the model could be an atom-cavity system with two atoms being confined in a single-mode optical microcavity~\cite{BJLiu2017}, a circuit-QED system with two superconducting qubits being coupled capacitively to a superconducting resonator~\cite{ZYXue2017}, or an ion-trap system with two trapped ions interacting with a single vibrational mode~\cite{Duan2001}, etc.
Within the rotating wave approximation, the interaction Hamiltonian of the qubit-resonator system is~($\hbar=1$):
\begin{eqnarray}\label{eq21}
H_{\rm I}&=&H_{\rm qr}+H_{\Omega}+H_{\Delta},\nonumber\\
H_{\rm qr}&=&g_1|0\rangle_1\langle e|a+g_2|1\rangle_2\langle e|a+\rm H.c.,\nonumber\\
H_{\Omega}&=&\frac{\Omega_1(t)}2|1\rangle_1\langle e|+\frac{\Omega_2(t)}2|0\rangle_2\langle e|+\rm H.c.,\nonumber\\
H_{\Delta}&=&-\sum_{j=1,2}\Delta|e\rangle_j\langle e|.
\end{eqnarray}
Here, for convenience, we ignore the phase factors of $\Omega_{1,2}(t)$ for the time being.
$H_{\rm qr}$ and $H_{\Omega}$ denote the qubit-resonator interaction and qubit-classical-field interaction, respectively. $H_{\Delta}$ means that all the interactions are with the same detuning $\Delta$. $a$ is the annihilation operator of the quantum-field mode in the resonator. We set a ket $|n\rangle_{\rm r}$~($n=0,1,2,...$) to mark the quantum-number state of the resonator, and the resonator is in $|0\rangle_{\rm r}$ initially.

As for quantum computation, the initial state of the qubit-resonator system is expanded by four states
\begin{eqnarray}\label{eq22}
|\phi_1\rangle&\equiv&|0\rangle_1|1\rangle_2|0\rangle_{\rm r},\quad|\phi_2\rangle\equiv|0\rangle_1|0\rangle_2|0\rangle_{\rm r},\nonumber\\
|\phi_6\rangle&\equiv&|1\rangle_1|1\rangle_2|0\rangle_{\rm r},\quad|\phi_7\rangle\equiv|1\rangle_1|0\rangle_2|0\rangle_{\rm r}.
\end{eqnarray}
After introducing the condition $g_{1,2}\gg\Omega_{1,2}(t)$, the evolution of the qubit-resonator system is banned for the initial states $|\phi_1\rangle$ and $|\phi_7\rangle$. For the initial states $|\phi_2\rangle$ or $|\phi_6\rangle$, the Hamiltonian~(\ref{eq21}) can be simplified into the following effective three-level Hamiltonian in the subspace \{$|\phi_2\rangle$, $|\Psi_0\rangle$, $|\phi_6\rangle$\}~(see Appendix for details)
\begin{eqnarray}\label{eq23}
H_{\rm I}^{\rm eff}=\frac{\hbar}{2}\left[
\begin{array}{ccc}
0                                &\Omega_p(t)                 &0                 \\
\Omega_p(t)^\ast                 &-2\Delta                    &\Omega_s(t)^\ast  \\
0                                &\Omega_s(t)                 &0                 \\
\end{array}
\right],
\end{eqnarray}
in which $\Omega_p(t)\equiv{g_1\Omega_2(t)}/{2G}$, $\Omega_s(t)\equiv-{g_2\Omega_1(t)}/{2G}$, and $|\Psi_0\rangle$ is defined in table~\ref{T4}. Here the phase factors of $\Omega_{1,2}(t)$ has been in consideration. The effective Hamiltonian~(\ref{eq23}) is of the exactly same form as the Hamiltonian~(\ref{eq1}). Naturally, by following the identical process of implementing speeded-up adiabatic single-qubit geometric gates in the section~\ref{S2}, the below two-qubit geometric gates in the computation space \{$|0\rangle_1|1\rangle_2$, $|0\rangle_1|0\rangle_2$, $|1\rangle_1|1\rangle_2$, $|1\rangle_1|0\rangle_2$\} could be implemented
\begin{eqnarray}\label{eq24}
U_2=
\left[
\begin{array}{cccc}
1&0                                                   &0                                            &0\\
0&\cos^2\eta+e^{i\gamma_\Phi}\sin^2\eta               &\cos\eta\sin\eta(e^{i\gamma_\Phi} -1)        &0\\
0&\cos\eta\sin\eta(e^{i\gamma_\Phi} -1)               &e^{i\gamma_\Phi}\cos^2\eta+\sin^2\eta        &0\\
0&0                                                   &0                                            &1\\
\end{array}
\right]\nonumber\\
\end{eqnarray}
for the cases of time-dependent detuning $\Delta=\Delta(t)$ and one-photon resonance $\Delta=0$;
\begin{eqnarray}\label{eq25}
U'_2=
\left[
\begin{array}{cccc}
1&0                                                   &0                                            &0\\
0&\cos\widetilde{\Gamma}+i\cos\widetilde{\Theta}\sin\widetilde{\Gamma}               &i\sin\widetilde{\Theta}\sin\widetilde{\Gamma}        &0\\
0&i\sin\widetilde{\Theta}\sin\widetilde{\Gamma}              &\cos\widetilde{\Gamma}-i\cos\widetilde{\Theta}\sin\widetilde{\Gamma}        &0\\
0&0                                                   &0                                            &1\\
\end{array}
\right]\nonumber\\
\end{eqnarray}
for the case of large detuning. Both $U_2$ and $U'_2$ are adequate to implement NTQG.

\begin{figure}[htb]\centering
\centering
\includegraphics[width=\linewidth]{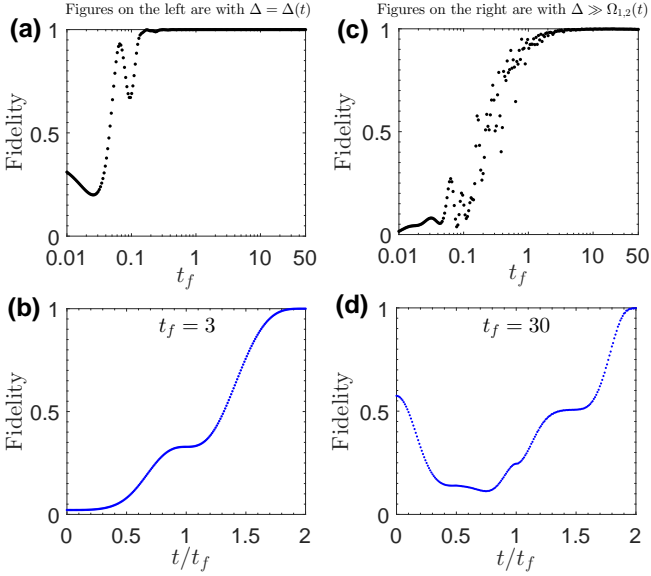}
\caption{Numerical demonstration of the speeded-up adiabatic NTQG~(\ref{eq24}) at time-dependent detuning $\Delta=\Delta(t)$ with the parameters in table~\ref{T1}~($\eta=\pi/2$) and $g_1=g_2=50$:
(a)~Effect of the value of $t_f$ on the final fidelity at the time $T=2t_f$;
(b)~Fidelity trends over time with $t_f=30$.
Numerical demonstration of the speeded-up adiabatic NTQG~(\ref{eq25}) at large detuning $\Delta=\Omega_{1,2}(t)$ with the parameters in table~\ref{T2} and $g_1=g_2=50$:
(c)~Effect of the value of $t_f$ on the final fidelity at the time $T=2t_f$;
(d)~Fidelity trends over time with $t_f=30$.}\label{f4}
\end{figure}
In Fig.~\ref{f4}, we show the numerical demonstration of the effectiveness for constructing the speeded-up adiabatic NTQG~(\ref{eq24})~[Figs.~\ref{f4}(a) and (b)] and (\ref{eq25})~[Figs.~\ref{f4}(c) and (d)]. The fidelity of the NTQG is defined as
$F(t)=|\langle{\Psi}_{U2}|\Psi(t)\rangle|^2$, where $|\Psi(t)\rangle$ is the state of the qubit-resonator system governed by the Sch\"{o}dinger equation based on the Hamiltonian~(\ref{eq21}). The initial state is $|\Psi(0)\rangle=\sin\alpha_1|\phi_1\rangle+\cos\alpha_1[e^{i\alpha_4}\sin\alpha_2|\phi_2\rangle+\cos\alpha_2(e^{i\alpha_5}\sin\alpha_3|\phi_6\rangle
+e^{i\alpha_6}\cos\alpha_3|\phi_7\rangle)]$ with $\alpha_1=\alpha_6=\pi/8$, $\alpha_2=\alpha_5=\pi/4$ and $\alpha_3=\alpha_4=3\pi/8$, without loss of generality. $|{\Psi}_{U2}\rangle$ is the target state after executing the NTQG $U_2$ or $U'_2$ on $|\Psi(0)\rangle$.

In order to ensure the condition $g_{1,2}\gg\Omega_{1,2}(t)$, we pick $g_1=g_2=50$ for all the three cases.
It is worth noting that the parameters in table~\ref{T3} for single-qubit $\sigma_z$ gate~(containing global phase) at large detuning can not be used for $U'_2$~(\ref{eq25}), because it corresponds to a trivial two-qubit gate. While those in table~\ref{T2} can be used for $U'_2$ in that it corresponds to a nontrivial two-qubit gate. Comparing Fig.~\ref{f4}(a) with (c), we know that a shorter operation time is required at time-dependent detuning than that at large detuning to make the fidelity up to unity.
It is because that only one limiting condition $g_{1,2}\gg\Omega_{1,2}(t)$ at time-dependent detuning but two conditions $g_{1,2}\gg\Omega_{1,2}(t)$ and $\Delta\gg\Omega_{p,s}^{\rm cd}(t)$ at large detuning need to be met. By the way, while the condition $\Omega_{\rm eff}^{\rm cd}(t)\gg\Delta_{\rm eff}(t)$ is also needed for the case of large detuning, it does not affect the required smallest $t_f$ but the largest, and Figs.~\ref{f4}(a) with (c) are both within it.
For a near-unity fidelity, the case of large detuning needs a longer operation time than the case of time-dependent detuning, as shown in Figs.~\ref{f4}(b) and (d). Anyhow Fig.~\ref{f4} shows the effectiveness of constructing speeded-up adiabatic NTQG at time-dependent detuning and large detuning. As a matter of fact, the case of one-photon resonance gives the same results as that of time-dependent detuning due to their equivalence property.

\section{Robustness against decay of the system}
\begin{figure}[htb]\centering
\centering
\includegraphics[width=\linewidth]{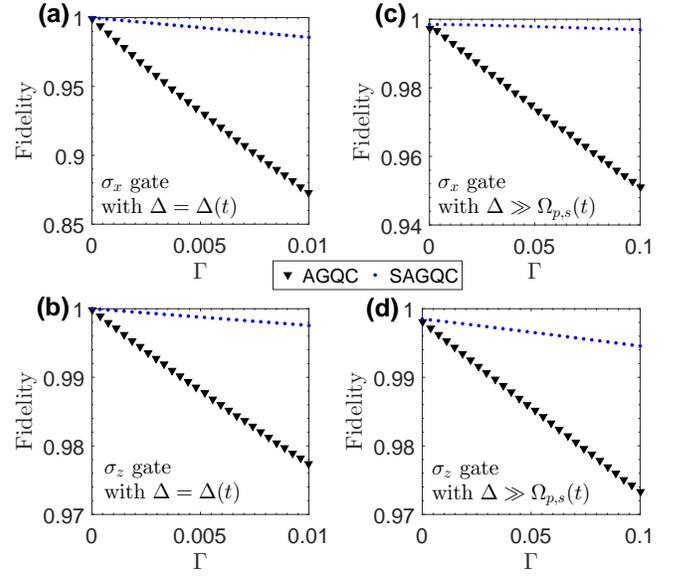}
\caption{Effect of the decay of the higher-energy level on the final fidelity for implementing USQG. The definition of fidelity is the same as that in Fig.~\ref{f2}.
(a) and (b)~Adiabatic implementation with $t_f=30$ and speeded-up adiabatic implementation of USQG in Eq.~(\ref{eq5}) with $t_f=3$, parameters in table~\ref{T1};
(c)~Adiabatic implementation with $t_f=5000$ and speeded-up adiabatic implementation of USQG in Eq.~(\ref{eq15}) with $t_f=30$, parameters in table~\ref{T2};
(d)~Adiabatic implementation with $t_f=2500$ and speeded-up adiabatic implementation of USQG in Eq.~(\ref{eq15}) with $t_f=30$, parameters in table~\ref{T3}.}\label{f5}
\end{figure}

In this section, we consider the effect of the decay of the system on the final fidelity to investigate the robustness of the proposals for implementing SAGQC. Taking the the decay of the system into account, the evolution of the system is dominated by the master equation under Markovian approximation
\begin{eqnarray}\label{eq26}
\dot{\rho}(t)&=&-\frac i\hbar[\mathcal{H},\rho(t)]+\sum_l[\mathcal{L}_l\rho(t){\mathcal{L}_l}^\dag\nonumber\\
&&-\frac12({\mathcal{L}_l}^\dag\mathcal{L}_l\rho(t)+\rho(t){\mathcal{L}_l}^\dag\mathcal{L}_l)],
\end{eqnarray}
with $\rho(t)$ denoting the density operator of the system, $\mathcal{H}$ the Hamiltonian~(\ref{eq1}) of the three-level system for USQG or (\ref{eq21}) of the qubit-resonator system for NTQG, and $l$ the number of Lindblad operators governing the decay of the system.
For the Hamiltonian~(\ref{eq1}), two Lindblad operators governing the decay of the three-level system~(i.e., the decay of the higher-energy level $|e\rangle$) are considered:
$\mathcal{L}_1=\sqrt{\Gamma_{0}}|0\rangle\langle e|$ and
$\mathcal{L}_2=\sqrt{\Gamma_{1}}|1\rangle\langle e|$,
in which $\Gamma_{m}$~($m=0,1$) is the decay rate from $|e\rangle$ to $|m\rangle$.
For the Hamiltonian~(\ref{eq21}), there are 5 Lindblad operators governing the the decay of the qubit-resonator system:
$\mathcal{L}_1=\sqrt{\Gamma_{1,0}}|0\rangle_1\langle e|$,
$\mathcal{L}_2=\sqrt{\Gamma_{1,1}}|1\rangle_1\langle e|$,
$\mathcal{L}_3=\sqrt{\Gamma_{2,0}}|0\rangle_2\langle e|$,
$\mathcal{L}_4=\sqrt{\Gamma_{2,1}}|1\rangle_2\langle e|$ and
$\mathcal{L}_5=\sqrt{\kappa}a$,
in which $\Gamma_{n,m}$~($n=1,2$) is the decay rate of Qubit~$n$ from $|e\rangle$ to $|m\rangle$, and $\kappa$ the decay rate of the resonator.
For simplicity, we set $\Gamma_{m}=\Gamma_{n,m}=\Gamma/2$ with $\Gamma$ being the total decay rate of a single qubit.

In Fig.~\ref{f5}, we plot the effect of the decay of the higher-energy level on the final fidelity to show the robustness of the proposals of implementing USQG. Apparently, with the increase of the decay rate, the final fidelity decreases linearly, with different ratios for different gates~($\sigma_x$ or $\sigma_z$) and different cases~(time-dependent detuning or large detuning). On one hand, with a fixed decay rate, in each subfigure the damage of the decay to the final fidelity of the adiabatic case is greater than that of the speeded-up adiabatic case for implementing USQG, besides the larger the decay rate, the greater the difference. The reason is that the adiabatic implementation of USQG requires longer operation time than the speeded-up case, which will definitely accumulate more decoherence induced by the decay and then lead to more damage to the final fidelity. On the other hand, contrasting the two cases, the damage of the decay to the final fidelity in the case of large detuning is significantly weaker than that in the case of time-dependent detuning, though the implementation of USQG in the case of large detuning requires longer operation time. It is because that in the case of large detuning, the state $|e\rangle$ is populated little~[the premise of the two-level effective Hamiltonian~(\ref{eq10})], while in the case of time-dependent detuning the state $|e\rangle$ plays an important role~[see Eqs.~(\ref{eq2}) and (\ref{eq6})] for the implementation of USQG in Eq.~(\ref{eq5}), which certainly will provide an environment where the decay could work well and then cause more damage to the final fidelity.

\begin{figure}[htb]\centering
\centering
\includegraphics[width=\linewidth]{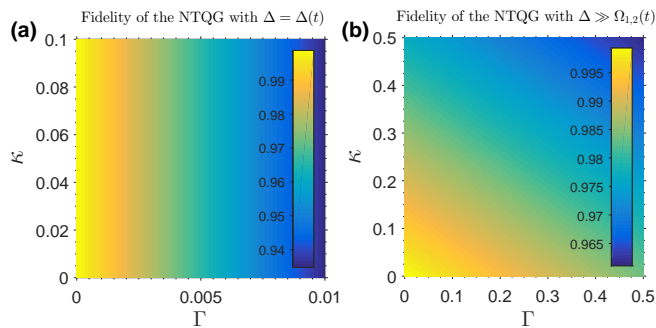}
\caption{Effect of the decay of the qubit-resonator system on the final fidelity for implementing NTQG. The definition of fidelity is the same as that in Fig.~\ref{f4}.
(a)~Speeded-up adiabatic implementation of NTQG in Eq.~(\ref{eq24}), parameters same as Fig.~\ref{f4}(b);
(b)~Speeded-up adiabatic implementation of NTQG in Eq.~(\ref{eq25}), parameters same as Fig.~\ref{f4}(d).}\label{f6}
\end{figure}
The effect of the decay of the qubit-resonator system on the final fidelity for implementing speeded-up adiabatic NTQG is shown in Fig.~\ref{f6}. During the implementation of the speeded-up adiabatic NTQG, we consider the condition $g_{1,2}\gg\Omega_{1,2}(t)$ that enables the effective three-level Hamiltonian~(\ref{eq23}). In fact, the condition $g_{1,2}\gg\Omega_{1,2}(t)$ makes the nonzero-number states of the resonator excluded in the evolution of the qubit-resonator system~(see the definitions of the states $|\phi_1\rangle$, $|\phi_2\rangle$, $|\Psi_0\rangle$, $|\phi_6\rangle$ and $|\phi_7\rangle$). Because the decay of the resonator comes from its nonzero-number states, the decay of the resonator is bound to affect little the final fidelity of implementing speeded-up adiabatic NTQG, which can be clearly demonstrated in Fig.~\ref{f6}(a). In Fig.~\ref{f6}(a) corresponding to the case of time-dependent detuning, the damage of the decay of the resonator to the final fidelity is invisible under the contrast with that of the decay of the two qubits. While in Fig.~\ref{f6}(b) corresponding to the case of large detuning, because the decay of the two qubits affects little the final fidelity yet, the damage of the decay of the resonator and that of the two qubits to the final fidelity are roughly equivalent and both slight.

According to the discussion above, we could conclude that in each case~(time-dependent detuning or large detuning), the implementation of SAGQC could not only shorten the operation time but also cut down the destructive effect of the decay of the system on the final fidelity. On the other hand, the destructive effect of the decay on the final fidelity in the case of time-dependent detuning is a little bit significant relatively, while that in the case of large detuning is slight. By the way, the robustness against the decay of the system in the unmentioned case of one-photon resonance is the same as that in the case of time-dependent detuning.

\section{Conclusion}
In conclusion, three coupling cases, i.e., time-dependent intermediate-level detuning, large intermediate-level detuning and one-photon resonance coupling are considered, respectively, to implement the universal SAGQC via counterdiabatic driving in $\Lambda$-type three-level system.

Different from other methods of STA, the counterdiabatic driving method makes the scheme has the following superiorities: (i) the property that the counterdiabatic Hamiltonian $H_{\rm cd}$ alone instead of $H_{0}+H_{\rm cd}$ is enough to speed up adiabatic evolution helps implement the universal SAGQC in the case of large detuning. (ii) this property even enables the implementation of the universal SAGQC in the case of one-photon resonance. Besides, the shortcoming of counterdiabatic driving is overcome in that no additional unaccessible coupling between two ground states is introduced but only the initial classical-field pulse shapes and phases are modified.

The discussion about the robustness against decay of the system is given. On one hand, the implementation of SAGQC is more robust than that of AGQC. On the other hand, the robustness in the case of large detuning is stronger than that in the cases of time-dependent detuning and one-photon resonance while at the cost of longer operation time. The work enriches the investigations of the universal geometric quantum computation in $\Lambda$-type three-level configuration, and may be used in the experiment of quantum computation in the future.

\section*{APPENDIX: Derivation of the effective Hamiltonian~(\ref{eq23})}
\begin{table}
\centering
\caption{\label{T4}Eigenvalues and eigenstates of the qubit-resonator interaction Hamiltonian $H_{\rm qr}$. Here we define $G=\sqrt{{g_1}^2+{g_2}^2}$.}
\footnotesize
\begin{tabular}{@{}ll}
\hline\hline
Eigenvalue&Eigenstate\\
\hline
$0$     &    $\{|\phi_2\rangle\}$, $\{|\phi_6\rangle\}$, $\{|\Psi_0\rangle\equiv(g_1|\phi_3\rangle-g_2|\phi_5\rangle)/G\}$\\
$\pm G$   &   $\{|\Psi_{\pm}\rangle\equiv(g_2|\phi_3\rangle\pm G|\phi_4\rangle+g_1|\phi_5\rangle)/\sqrt2G\}$ \\
\hline\hline
\end{tabular}
\end{table}
Obviously, the evolution of the qubit-resonator system is banned when the initial state is $|\phi_1\rangle=|0\rangle_1|1\rangle_2|0\rangle_{\rm r}$, because $|\phi_1\rangle$ is decoupled to the Hmailtonian~(\ref{eq21}).

When the initial state is $|\phi_2\rangle=|0\rangle_1|0\rangle_2|0\rangle_{\rm r}$ or $|\phi_6\rangle=|1\rangle_1|1\rangle_2|0\rangle_{\rm r}$,
the qubit-resonator system will evolve in the following Hilbert subspace
\begin{align*}\label{eA1}
|\phi_2\rangle&=|0\rangle_1|0\rangle_2|0\rangle_{\rm r},\quad|\phi_3\rangle=|0\rangle_1|e\rangle_2|0\rangle_{\rm r},\nonumber\\
|\phi_4\rangle&=|0\rangle_1|1\rangle_2|1\rangle_{\rm r},\quad|\phi_5\rangle=|e\rangle_1|1\rangle_2|0\rangle_{\rm r},\nonumber\\
|\phi_6\rangle&=|1\rangle_1|1\rangle_2|0\rangle_{\rm r}.
\tag*{(A1)}
\end{align*}
With the basis states in Eq.~\ref{eA1}, the eigenvalues and eigenstates of the qubit-resonator interaction Hamiltonian $H_{\rm qr}$ can be calculated out and then listed in table~\ref{T4}. Now with the eigenstates of $H_{\rm qr}$ being the basis states, the qubit-resonator interaction Hamiltonian in diagonalization represented by its eigenstates becomes $H'_{\rm qr}=G(|\Psi_+\rangle\langle\Psi_+|-|\Psi_-\rangle\langle\Psi_-|)$. The Hamiltonian~(\ref{eq21}) becomes
\begin{align*}\label{eA2}
H'_{\rm I}&=H'_{\rm qr}+H'_{\Omega}+H'_{\Delta},\nonumber\\
H'_{\Omega}&=\frac{\Omega_2(t)}{2G}|\phi_2\rangle[g_1\langle\Psi_0|+\frac{g_2}{\sqrt2}(\langle\Psi_+|+\langle\Psi_-|)]+\frac{\Omega_1(t)}{2G}\nonumber\\
&\quad\times[\frac{g_1}{\sqrt2}(|\Psi_+\rangle+|\Psi_-\rangle)-g_2|\Psi_0\rangle]\langle\phi_6|+{\rm H.c.},\nonumber\\
H'_{\Delta}&=-\Delta[|\Psi_0\rangle\langle\Psi_0|+\frac12(|\Psi_+\rangle+|\Psi_-\rangle)(\langle\Psi_+|+\langle\Psi_-|)].
\tag*{(A2)}
\end{align*}
Through the unitary transformation $\exp(-iH'_{\rm qr}t)$, all transitions involving nonzero-eigenvalue eigenstates $|\Psi_\pm\rangle$ are of high-frequency oscillations by considering the condition $g_{1,2}\gg\Omega_{1,2}(t)$. Therefore, after neglecting high-frequency oscillation terms, we could obtain an effective Hamiltonian solely involving the zero-eigenvalue eigenstates of $H_{\rm qr}$
\begin{align*}\label{eA3}
H'_{\rm eff}&=[\frac{g_1\Omega_2(t)}{2G}|\phi_2\rangle\langle\Psi_0|-\frac{g_2\Omega_1(t)}{2G}|\phi_6\rangle\langle\Psi_0|+{\rm H.c.}]\nonumber\\
&\quad-\Delta|\Psi_0\rangle\langle\Psi_0|.
\tag*{(A3)}
\end{align*}

When the initial state is $|\phi_7\rangle=|1\rangle_1|0\rangle_2|0\rangle_{\rm r}$ the qubit-resonator system will evolve in the following Hilbert subspace
\begin{align*}\label{eA4}
&|\phi_7\rangle=|1\rangle_1|0\rangle_2|0\rangle_{\rm r},\quad|\phi_8\rangle=|e\rangle_1|0\rangle_2|0\rangle_{\rm r},\nonumber\\
&|\phi_9\rangle=|1\rangle_1|e\rangle_2|0\rangle_{\rm r},\quad|\phi_{10}\rangle=|0\rangle_1|0\rangle_2|1\rangle_{\rm r},\nonumber\\
&|\phi_{11}\rangle=|e\rangle_1|e\rangle_2|0\rangle_{\rm r},\quad|\phi_{12}\rangle=|1\rangle_1|1\rangle_2|1\rangle_{\rm r},\nonumber\\
&|\phi_{13}\rangle=|0\rangle_1|e\rangle_2|1\rangle_{\rm r},\quad|\phi_{14}\rangle=|e\rangle_1|1\rangle_2|1\rangle_{\rm r},\nonumber\\
&|\phi_{15}\rangle=|0\rangle_1|1\rangle_2|2\rangle_{\rm r}.
\tag*{(A4)}
\end{align*}
Similar to the case of the initial state being $|\phi_2\rangle$ or $|\phi_6\rangle$, consider the condition $g_{1,2}\gg\Omega_{1,2}(t)$, and then the Hamiltonian~(\ref{eq21}) could be simplified into an effective Hamiltonian only involving zero-eigenvalue eigenstates of $H_{\rm qr}$. In the case of the initial state being $|\phi_7\rangle$, $|\phi_7\rangle$ is the unique zero-eigenvalue eigenstate of $H_{\rm qr}$. Therefore, we can say that the evolution of the qubit-resonator system with $|\phi_7\rangle$ being the initial state is banned.

\section*{ACKNOWLEDGEMENTS}
This  work  was  supported  by China Postdoctoral Science Foundation under Grant No. 2018T110735 and National Natural Science Foundation of China under No. 11804308.

\end{document}